\documentclass[12pt]{article}

\usepackage{lscape}
\usepackage{pdflscape}
\usepackage[figuresright]{rotfloat}[2004/01/04]

\usepackage{graphicx}
\usepackage{caption}
\usepackage{float}
\usepackage{booktabs,tabularx}
\usepackage{textcomp}
\usepackage{makecell}
\usepackage[semicolon,authoryear]{natbib}
\usepackage{rotating}
\usepackage[table]{xcolor}

\usepackage{dsfont}

\usepackage{array}
\newcolumntype{L}[1]{>{\raggedright\let\newline\\\arraybackslash\hspace{0pt}}m{#1}}
\newcolumntype{C}[1]{>{\centering\let\newline\\\arraybackslash\hspace{0pt}}m{#1}}
\newcolumntype{R}[1]{>{\raggedleft\let\newline\\\arraybackslash\hspace{0pt}}m{#1}}

\usepackage{tikz}
\usetikzlibrary{arrows.meta}
\usetikzlibrary{shapes,snakes}
\restylefloat{table}

\usepackage{multirow}

\usepackage{amsmath}
\usepackage{mathtools}
\usepackage{mathrsfs}
\usepackage{amssymb}
\usepackage{flexisym}
\usepackage{bm}
\usepackage{mathptmx}
\usepackage[nolists,nomarkers]{endfloat}
\usepackage{epsfig}
\usepackage[margin=1in]{geometry}
\usepackage{authblk}
\usepackage{scalerel,stackengine}
\stackMath
\newcommand\reallywidehat[1]{%
	\savestack{\tmpbox}{\stretchto{%
			\scaleto{%
				\scalerel*[\widthof{\ensuremath{#1}}]{\kern-.6pt\bigwedge\kern-.6pt}%
				{\rule[-\textheight/2]{1ex}{\textheight}}%WIDTH-LIMITED BIG WEDGE
			}{\textheight}% 
		}{0.5ex}}%
	\stackon[1pt]{#1}{\tmpbox}%
}
\DeclareMathOperator*{\argmin}{arg\,min}

\floatstyle{plaintop}
\floatname{table}{Table}
\newfloat{table}{tbhp}{lop}[section]
\floatname{figure}{Figure}
\newfloat{figure}{h}{lof}[section]

\usepackage{xr}
\makeatletter
\newcommand*{\addFileDependency}[1]{% argument=file name and extension
	\typeout{(#1)}
	\@addtofilelist{#1}
	\IfFileExists{#1}{}{\typeout{No file #1.}}
}
\makeatother

\newcommand*{\myexternaldocument}[1]{%
	\externaldocument{#1}%
	\addFileDependency{#1.tex}%
	\addFileDependency{#1.aux}%
}

\myexternaldocument{supp}
\begin{document}

\setlength{\textheight}{625pt} \setlength{\baselineskip}{26pt}	
\title{Power Prior Models for Treatment Effect Estimation in a Small n, Sequential, Multiple Assignment, Randomized Trial}
\author[1]{Yan-Cheng Chao}
\author[1]{Thomas M. Braun}
\author[2]{Roy N. Tamura}
\author[1]{Kelley M. Kidwell}
\affil[1]{\footnotesize Department of Biostatistics, School of Public Health, University of Michigan, Ann Arbor, MI 48109, U.S.A.}
\affil[2]{\footnotesize Health Informatics Institute, University of South Florida, Tampa, FL 33620, U.S.A.}
\date{}
\maketitle

\begin{abstract} 
	A small n, sequential, multiple assignment, randomized trial (snSMART) is a small sample, two-stage design where participants receive up to two treatments sequentially, but the second treatment depends on response to the first treatment. The treatment effect of interest in an snSMART is the first-stage response rate, but outcomes from both stages can be used to obtain more information from a small sample. A novel way to incorporate the outcomes from both stages applies power prior models, in which first stage outcomes from an snSMART are regarded as the primary data and second stage outcomes are regarded as supplemental. We apply existing power prior models to snSMART data, and we also develop new extensions of power prior models. All methods are compared to each other and to the Bayesian joint stage model (BJSM) via simulation studies. By comparing the biases and the efficiency of the response rate estimates among all proposed power prior methods, we suggest application of Fisher's exact test or the Bhattacharyya's overlap measure to an snSMART to estimate the treatment effect in an snSMART, which both have performance mostly as good or better than the BJSM. We describe the situations where each of these suggested approaches is preferred. 
	
	\noindent
	
	\vspace{0.25in}
	
\end{abstract}

\newpage
\section{Introduction}
\label{sec:intro}

In rare disease studies, estimating treatment effects efficiently is often a challenging task because information is collected from a relatively small number of participants. Developed to meet this challenge, a small n, sequential, multiple assignment, randomized trial (snSMART) is a two-stage design where participants are given up to two treatments sequentially; whether they receive the same or different treatment in the second stage depends on how they respond to the first stage treatment \citep{tamura2016small}. Primary interest in an snSMART is the first stage treatment effect, but when multiple outcomes are obtained from each participant, a method to combine the information across stages can be used to efficiently estimate the treatment effects of interest.

Frequentist and Bayesian approaches have been proposed to pool the results together for estimation. \cite{tamura2016small} presented a weighted Z-statistic to perform the estimation, but the Z-statistic is not based on all the collected data. To address these limitations, \cite{wei2018bayesian} and \cite{chao2020dynamic} presented both a Bayesian joint stage model (BJSM) and a joint stage regression model, each of which includes parameters that link first and second stage treatment responses to provide more efficient treatment effect estimates. Here, we present an alternative approach that links data from the two stages through a power prior, which was first proposed by \cite{ibrahim2000power}. 

A power prior contains the likelihood of the historical data, power parameters that quantify the compatibility of the historical and the current data, and prior distributions for the parameters in the likelihood of the current data. The power parameters can be either fixed or random and there are numerous ways the parameters are specified or determined. Extensions of this power prior approach include modified power priors, or normalized power priors \citep{duan2006evaluating,neuenschwander2009note,hobbs2011hierarchical,banbeta2019modified,van2018including}, power prior in Bayesian hierarchical models \citep{chen2006relationship}, commensurate power priors \citep{hobbs2011hierarchical,van2018including}, power priors with an empirical Bayesian approach \citep{gravestock2017adaptive} and power priors with a likelihood-based weight selection criterion \citep{ibrahim2003optimality,ibrahim2015power}.

\cite{pan2017calibrated} proposed a calibrated power prior that utilizes a nonparametric Kolmogorov-Smirnov statistic to measure the compatibility of historical and current data in biosimilar designs. \cite{nikolakopoulos2018dynamic} developed another calibrated power prior that quantifies the conflict of historical to current data through prior predictive p-values. \cite{li2020pa} applied the notion of a power prior model to control information borrowing through Bayesian model averaging between pediatric and adult phase I oncology trials.

In previous studies, the idea of power prior models was applied to control how much information should be borrowed from historical data or earlier trials to a current trial. However, information sharing is also crucial in a multistage clinical trial, which motivates our work. In this study, we propose a novel application of power prior models to the estimation of treatment effects in an snSMART, which is a two-stage design. In addition, we first introduce novel measures of closeness to describe the compatibility of stage 1 and 2 data in our snSMART. In our setting, we consider stage 1 responses as ``current" data and stage 2 responses as ``historical" data, which may seem counterintuitive. However, because a second stage outcome is obtained after a first stage outcome, second stage outcomes are conditional on the treatments received in the first stage and response to that first stage treatment. Because of this biased sampling scheme, the second stage outcomes are viewed as supplemental data, and the first stage outcomes are viewed as the primary data, since they are collected in an unbiased, randomized design.

Small sample size is another challenge when applying power prior models to the snSMART setting. In existing designs, the historical data are often assumed to come from a multitude of participants who received the same treatment. In contrast, in an snSMART, it is possible that outcomes will only be obtained from a very small number of participants in the second stage. The operating characteristics of power prior models with small samples has not been investigated before, and thus, we seek to examine their performance in the snSMART setting relative to the existing BJSM.

In our current work, we propose three different power prior models to estimate the response rates of three active treatments in an snSMART. In Section \ref{sec:example}, we motivate the use of power prior models in snSMART designs and briefly describe the existing BJSM. In Section \ref{sec:method}, we present the power prior models with different power parameter specification approaches. In Section \ref{sec:simulation}, we use simulations to examine how these power prior models perform and compare them to the BJSM under different scenarios, and we close with a discussion in Section \ref{sec:discussion}. 

\section{Motivating example and existing methods}
\label{sec:example}
\subsection{ARAMIS trial}
\label{sec:ARAMIS}
\begin{figure}[H]
	\centering
	\includegraphics[width=\textwidth]{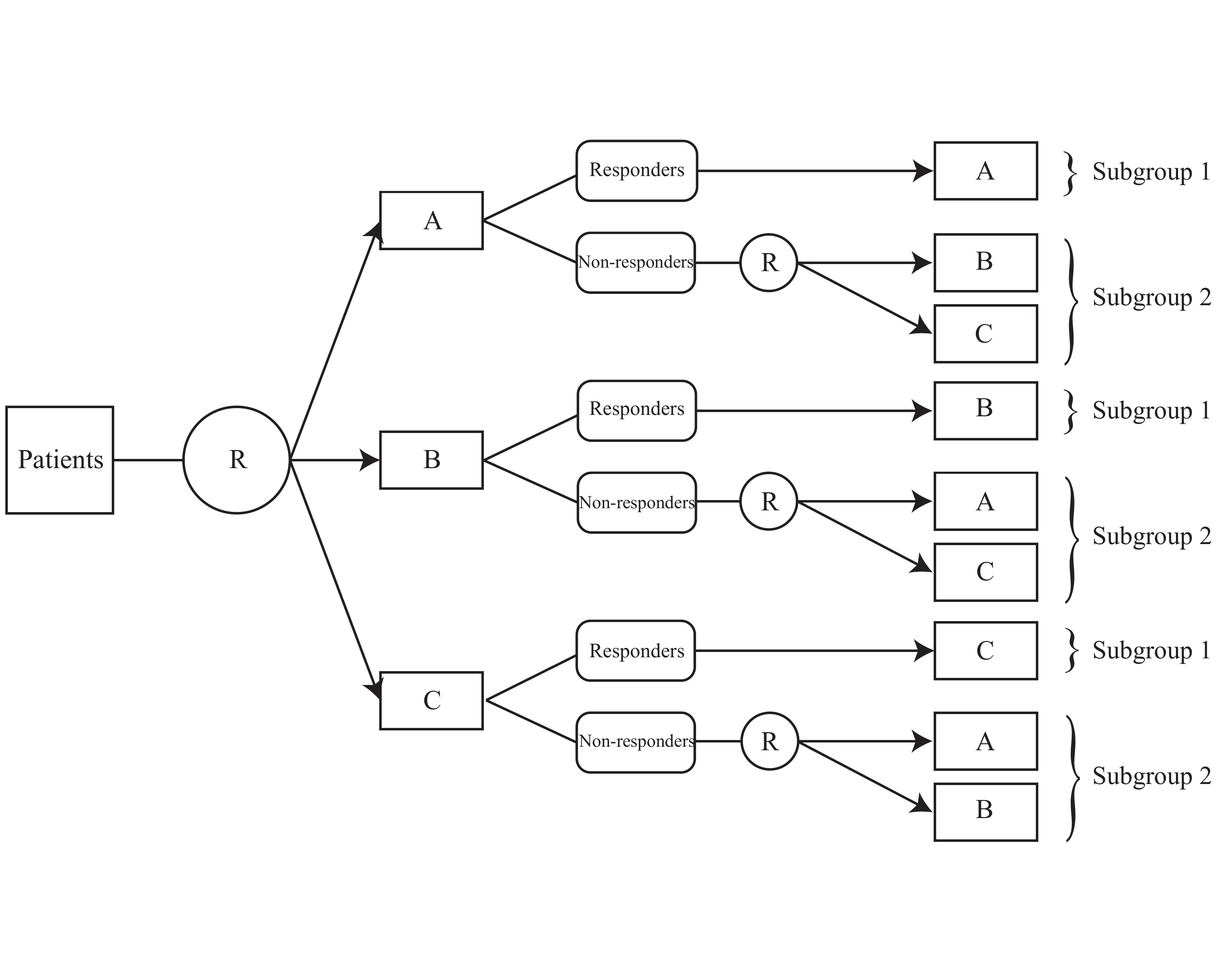}
	\caption{A small n, sequential, multiple assignment, randomized trial design in our study. R represents randomization to the following treatments. Subgroups refer to subsets of second stage data applied in a power prior model.}
	\label{fig:design}
\end{figure}

Our methods are motivated by the snSMART, A RAndomized Multicenter study for Isolated Skin vasculitis (ARAMIS) \citep{micheletti2020protocol}, \cite{wei2018bayesian} and \cite{chao2020dynamic} and shown in Figure \ref{fig:design}. In brief, all enrolled individuals are randomized to one of the three treatments in the first stage. During a specific period of follow-up of six months, each individual is assessed for a response. The individuals who respond in the first stage receive the same treatment in the second stage, while non-responders in the first stage are randomized to one of the alternative treatments in the second stage and followed for six more months for response.

The first stage is a traditional randomized trial; thus, we can estimate treatment effects using only the first stage data. In the proposed power prior methods, these first stage outcomes are called ``current data". By contrast, the second stage outcomes alone could not be used to correctly estimate the response rates because they are conditional on first stage treatment and responses to that treatment. Thus, second stage outcomes serve as ``historical" data. Inclusion of ``historical" data can provide additional information and increase the efficiency of estimation of treatment effects in small samples. Thus, the application of power prior models to our setting provides a way to incorporate both stages of data such that first stage data are weighted fully, and second stage data receive partial weight through the power prior to provide more efficient treatment estimates in small samples.

\subsection{Joint stage models}
\label{sec:JSM}

Frequentist and Bayesian joint stage models are existing approaches that estimate the treatment effects in an snSMART, where the details can be found in \cite{wei2018bayesian} and \cite{chao2020dynamic}. Because the results from both models are similar, we briefly present the BJSM here due to our focus on Bayesian methods.

The (first stage) response rate of a treatment $k$ is denoted by $\pi_k$, where $k=A, B, C$. Since the response rate of a treatment in the second stage can differ from that in the first stage, and because stage 2 response rates are conditional on stage 1 treatments and responses, we denote the second stage response rates of the first stage responders to treatment $k$ by $\beta_{1}\pi_k$, and the second stage response rates of the first stage non-responders to $k$ who receive $k'$ in the second stage by $\beta_{0}\pi_{k'}$. $\beta_{1}$ and $\beta_{0}$ are called linkage parameters for stage 1 responders and non-responders, respectively, because they link the first stage and second stage response rates. An assumption of the BJSM is that the linkage parameters, $\beta_0$ and $\beta_1$, do not depend on the first and second stage treatments received. The parameters, $\pi_k$, $\beta_{1}$ and $\beta_{0}$, can be estimated via Markov Chain Monte Carlo with appropriate prior distributions on these parameters.

However, we may not have \textit{a priori} information about the possible relationship between first stage and second response rates, particularly in the rare disease settings, which may make it difficult to pre-specify prior distributions of the linkage parameters. Thus, the power prior approaches presented next provide a framework to circumvent the requirement of assuming the proportionality of response rates from the stage 1 to 2.

%
%Also for a subject $i=1,\dotsc,N$, where $N$ is the sample size of an snSMART, we let $Y_{i1}^k$ and $Y_{i2}^{k'}$ be binary responses from a subject $i$ who receives treatment $k$ in the first stage and $k'$ in the second stage, respectively. 
%
%The BJSM is formulated as follows.
%
%\begin{equation}
%\label{eq:BJSM_Y1}
%Y_{i1}^k|\pi_k\sim\text{Bernoulli}(\pi_k)
%\end{equation}
%\begin{equation}
%\label{eq:BJSM_Y2}
%Y_{i2}^{k'}|Y_{i1}^k,\pi_k,\pi_{k'},\beta_{0k},\beta_{1k}\sim\text{Bernoulli}((\beta_{1k}\pi_k)^{Y_{i1}^k}(\beta_{0k}\pi_{k'})^{1-Y_{i1}^k})
%\end{equation}
%
%where the prior distributions of $\pi_k$, $\beta_{0k}$ and $\beta_{1k}$ are beta(0.4, 1.6), beta(1.6,0.4) and Pareto(1,3), respectively. The prior distributions can change based on the assumptions on these parameters. The detailed information about this model can be found in \cite{wei2018bayesian}.

\section{Methods}
\label{sec:method}

We first briefly review the power prior models and their associated notation. We let $\boldsymbol{\pi}=\{\pi_A,\pi_B,\pi_C\}$, where the elements are the response rates of treatments A, B, and C, respectively, and $\boldsymbol{\delta}=\{\delta_j\}$ denote power parameters for different subgroups of individuals, where $j=1,\dotsc, J$ and $J$ is the number of subgroups. In our design, we separate the second stage data into two distinct sets: those from first stage responders and those from first stage non-responders. The individuals in these two subgroups are assumed to share some common within-group characteristics that may affect how they respond to the second stage treatments. Thus, each subgroup can be regarded as a distinct set of ``historical" data, and we assume that $J=2$ in this study. We also made this assumption of $J=2$ because a parsimonious model is preferred when the sample size is small, and two power parameters mimics the two linkage parameters from the BJSM. Let $n_k^{(1)}$ and $Z_k^{(1)}$ denote the number of individuals assigned to treatment $k$ and the corresponding number of responders to $k$ in stage 1, respectively, where $k=A,B,C$. Similarly, we let $n_{k,j}^{(2)}$ and $Z_{k,j}^{(2)}$ be the numbers of individuals in stage 2 assigned to treatment $k$ within subgroup $j$ and the corresponding number of responders to $k$ in subgroup $j$, respectively. Let $\mathbf{D^{(1)}}=\{n_k^{(1)}, Z_k^{(1)}; k=A,B,C\}$ and $\mathbf{D^{(2)}}=\{n_{k,j}^{(2)}, Z_{k,j}^{(2)}; k=A,B,C; j=1,\dotsc, J\}$.

In its simplest form, the joint power prior distribution of the first stage response rates in our setting can be formulated as

%\begin{equation}
%\label{eq:gen_power_prior}
%p(\boldsymbol{\pi}|\mathbf{D^{(2)}},\boldsymbol{\delta})\propto \prod_{k=A,B,C}\left[\prod_{j\in\Omega_k} L(Z_{k,j}^{(2)}; \pi_k)^{\delta_j}\right]p_0(\pi_k)
%\end{equation}

\begin{equation}
\label{eq:gen_power_prior}
p(\boldsymbol{\pi}|\mathbf{D^{(2)}},\boldsymbol{\delta})\propto \prod_{k=A,B,C}\left[\prod_{j=1,\dotsc, J} L(Z_{k,j}^{(2)}; \pi_k)^{\delta_j}\right]p_0(\pi_k)
\end{equation}

where $L(Z_{k,j}^{(2)}; \pi_k)$ is a likelihood function for second stage outcomes, $p_0(\pi_k)$ is the initial prior for $\pi_k$, and $0\leq \delta_j\leq 1$ for all $j$. We interpret $\delta_j$ as a measure of compatibility of  the ``current" data and the ``historical" data from subgroup $j$. When $\delta_j=0$, the corresponding ``historical" data, i.e., second stage data, from subgroup $j$ contribute nothing to the estimation of response rates, while $\delta_j=1$ indicates that the corresponding ``historical" data from subgroup $j$ can be pooled together with ``current" data. When combining with the likelihood function of first stage outcomes, the posterior distribution of $\boldsymbol{\pi}$ is

%\begin{align}
%\label{eq:gen_posterior}
%q(\boldsymbol{\pi}|\mathbf{D^{(1)}}, \mathbf{D^{(2)}},\boldsymbol{\delta})
%&\propto \left[\prod_{k=A,B,C} L(Z_k^{(1)}; \pi_k)\right]p(\boldsymbol{\pi}|\mathbf{D^{(2)}},\boldsymbol{\delta}) \nonumber\\
%&=\prod_{k=A,B,C}\left[L(Z_k^{(1)}; \pi_k)\prod_{j\in\Omega_k} L(Z_{k,j}^{(2)}; \pi_k)^{\delta_j}\right]p_0(\pi_k)
%\end{align}

\begin{align}
\label{eq:gen_posterior}
q(\boldsymbol{\pi}|\mathbf{D^{(1)}}, \mathbf{D^{(2)}},\boldsymbol{\delta})
&\propto \left[\prod_{k=A,B,C} L(Z_k^{(1)}; \pi_k)\right]p(\boldsymbol{\pi}|\mathbf{D^{(2)}},\boldsymbol{\delta}) \nonumber\\
&=\prod_{k=A,B,C}\left[L(Z_k^{(1)}; \pi_k)\prod_{j=1,\dotsc, J} L(Z_{k,j}^{(2)}; \pi_k)^{\delta_j}\right]p_0(\pi_k)
\end{align}

The key issue in the application of power prior models lies in the choice of $\delta_j$. Thus, we next introduce three types of approaches for choosing $\delta_j$ and investigate to what extent stage 2 data can be incorporated with stage 1 data to estimate $\pi_k$.

\subsection{Power prior models with likelihood-type criteria}
\label{sec:PLC_MLC}
The power parameters $\delta_1$ and $\delta_2$ can be taken as fixed values and determined by likelihood-type criteria, which was first proposed by \cite{ibrahim2003optimality} and extended from Bayesian Information Criterion (BIC). The rationale of utilizing likelihood-type criteria is to use both ``current" and ``historical" data to choose the optimal values for $\delta_1$ and $\delta_2$ that minimize the criteria function. Two criteria applied to power prior models are the penalized likelihood-type criterion (PLC) \citep{ibrahim2003optimality,ibrahim2015power} and the marginal likelihood criterion (MLC) \citep{ibrahim2015power,gravestock2017adaptive}, the latter of which is also referred to as the empirical Bayesian method. 

For the PLC, the ``current" and ``historical" data are combined in the function
%\begin{align}
%\label{eq:m_star}
%m^*(\boldsymbol{\delta})&=\int_{\boldsymbol{\pi}}\prod_k \left[L(Z_k^{(1)};\pi_k)\prod_{j\in\Omega_k}L(Z_{k,j}^{(2)};\pi_k)^{\delta_j}p_0(\pi_k)\right]d\boldsymbol{\pi}  \nonumber \\
%&=M\prod_{k}\left\{B\left(Z_k^{(1)}+\sum_{j\in\Omega_k}Z_{k,j}^{(2)}\delta_j +a_{\pi},n_k^{(1)}-Z_k^{(1)}+\sum_{j\in\Omega_k}\left(n_{k,j}^{(2)}-Z_{k,j}^{(2)}\right)\delta_j+b_{\pi}\right)
%\right\}
%\end{align}

\begin{align}
\label{eq:m_star}
m^*(\boldsymbol{\delta})&=\int_{\boldsymbol{\pi}}\prod_k \left[L(Z_k^{(1)};\pi_k)\prod_{j}L(Z_{k,j}^{(2)};\pi_k)^{\delta_j}p_0(\pi_k)\right]d\boldsymbol{\pi}  \nonumber \\
&=M\prod_{k}\left\{B\left(Z_k^{(1)}+\sum_{j}Z_{k,j}^{(2)}\delta_j +a_{\pi},n_k^{(1)}-Z_k^{(1)}+\sum_{j}\left(n_{k,j}^{(2)}-Z_{k,j}^{(2)}\right)\delta_j+b_{\pi}\right)
\right\}
\end{align}

\noindent where $M$ is a constant unrelated to any of the parameters, and $B(\cdot,\cdot)$ is a beta function. The power parameters $\delta_1$ and $\delta_2$ can then be determined by minimizing the PLC function 

\begin{equation}
\label{eq:G_function}
G(\boldsymbol{\delta})=-2\log\left[m^*(\boldsymbol{\delta})\right]+\sum_j\frac{\log(\sum_k n_{k,j}^{(2)})}{\delta_j}.
\end{equation}

\noindent The penalty term $\sum_j  [\log(\sum_k n_{k,j}^{(2)})/\delta_j]$ allows for the chosen $\delta_j$ being higher when the sample size of subgroup $j$ is larger, which corresponds to more weight applied to a subgroup with a larger sample size. After the optimal $\boldsymbol{\delta}$ is determined by $\boldsymbol{\delta}^{PLC}=\argmin_{\boldsymbol{\delta}} G(\boldsymbol{\delta})$, we then treat $\boldsymbol{\delta}^{PLC}$ as fixed and use Equation (\ref{eq:gen_posterior}) to obtain the posterior distribution of all $\pi_k$. 

For the MLC, we use the marginal likelihood of $\boldsymbol{\delta}$

\begin{align}
m(\boldsymbol{\delta})
&=\frac{\int_{\boldsymbol{\pi}}\prod_k \left[L(Z_k^{(1)};\pi_k)\prod_{j}L(Z_{k,j}^{(2)};\pi_k)^{\delta_j}p_0(\pi_k)\right]d\boldsymbol{\pi}}{\int_{\boldsymbol{\pi}}\prod_k \left[\prod_{j}L(Z_{k,j}^{(2)};\pi_k)^{\delta_j}p_0(\pi_k)\right]d\boldsymbol{\pi}} \nonumber \\
&=M'\frac{\prod_k\left\{B\left(Z_k^{(1)}+\sum_{j}Z_{k,j}^{(2)}\delta_j +a_{\pi},n_k^{(1)}-Z_k^{(1)}+\sum_{j}\left(n_{k,j}^{(2)}-Z_{k,j}^{(2)}\right)\delta_j+b_{\pi}\right)\right\}}
{\prod_k\left\{B\left(\sum_{j}Z_{k,j}^{(2)}\delta_j +a_{\pi},\sum_{j}\left(n_{k,j}^{(2)}-Z_{k,j}^{(2)}\right)\delta_j+b_{\pi}\right)\right\} }
\end{align}

\noindent where $M'$ is a constant unrelated to any of the parameters. Values for the power parameters are determined as $\boldsymbol{\delta}^{MLC}=\argmin_{\boldsymbol{\delta}}\{-2\log[m(\boldsymbol{\delta})]\}$.

%\begin{align}
%m(\boldsymbol{\delta})
%&=\frac{\int_{\boldsymbol{\pi}}\prod_k \left[L(Z_k^{(1)};\pi_k)\prod_{j\in\Omega_k}L(Z_{k,j}^{(2)};\pi_k)^{\delta_j}p_0(\pi_k)\right]d\boldsymbol{\pi}}{\int_{\boldsymbol{\pi}}\prod_k \left[\prod_{j\in\Omega_k}L(Z_{k,j}^{(2)};\pi_k)^{\delta_j}p_0(\pi_k)\right]d\boldsymbol{\pi}} \nonumber \\
%&=M'\frac{\prod_k\left\{B\left(Z_k^{(1)}+\sum_{j\in\Omega_k}Z_{k,j}^{(2)}\delta_j +a_{\pi},n_k^{(1)}-Z_k^{(1)}+\sum_{j\in\Omega_k}\left(n_{k,j}^{(2)}-Z_{k,j}^{(2)}\right)\delta_j+b_{\pi}\right)\right\}}
%{\prod_k\left\{B\left(\sum_{j\in\Omega_k}Z_{k,j}^{(2)}\delta_j +a_{\pi},\sum_{j\in\Omega_k}\left(n_{k,j}^{(2)}-Z_{k,j}^{(2)}\right)\delta_j+b_{\pi}\right)\right\} }
%\end{align}

\subsection{Modified power prior model}
\label{sec:modified_prior}

The modified power prior (MPP) model proposed by \cite{duan2006evaluating} treats $\delta_1$ and $\delta_2$ as random variables; \cite{banbeta2019modified} applied the MPP to estimate treatment effects that incorporate control arms into a current trial. In our study, the MPP is given by

%\begin{equation}
%\label{eq:MPP}
%p_{MPP}(\boldsymbol{\pi},\boldsymbol{\delta}|\mathbf{D^{(2)}})\propto \frac{\left[\prod_{k}\prod_{j\in\Omega_k} L(Z_{k,j}^{(2)};\pi_k)^{\delta_j}\right]\left[\prod_{j}p_0(\delta_j)\right]\left[\prod_{k}p_0(\pi_k)\right]}{C(\boldsymbol{\delta})}
%\end{equation}

\begin{equation}
\label{eq:MPP}
p_{MPP}(\boldsymbol{\pi},\boldsymbol{\delta}|\mathbf{D^{(2)}})= \frac{\left[\prod_{k}\prod_{j} L(Z_{k,j}^{(2)};\pi_k)^{\delta_j}\right]\left[\prod_{j}p_0(\delta_j)\right]\left[\prod_{k}p_0(\pi_k)\right]}{C(\boldsymbol{\delta})}
\end{equation}

where

%\begin{equation}
%\label{eq:norm_constant}
%C(\boldsymbol{\delta})=\int_{\boldsymbol{\pi}}\left[\prod_{k}\prod_{j\in\Omega_k} L(Z_{k,j}^{(2)};\pi_k)^{\delta_j}\right]\left[\prod_{k}p_0(\pi_k)\right]d\boldsymbol{\pi}
%\end{equation}

\begin{equation}
\label{eq:norm_constant}
C(\boldsymbol{\delta})=\int_{\boldsymbol{\pi}}\left[\prod_{k}\prod_{j} L(Z_{k,j}^{(2)};\pi_k)^{\delta_j}\right]\left[\prod_{k}p_0(\pi_k)\right]d\boldsymbol{\pi}
\end{equation}

\noindent and $p_0(\delta_j)$ is an initial prior distribution of $\delta_j$. The normalizing constant $C(\boldsymbol{\delta})$ is necessary in the formulation of MPP when $\delta_j$ is random to enforce the likelihood principle \citep{duan2006evaluating,banbeta2019modified}. 

%In other word, if the likelihood principle is obeyed, multiplying a likelihood function by a constant will not change the inference on the parameters. However, if the normalizing constant $C(\boldsymbol{\delta})$ does not exist in Equation (\ref{eq:MPP}), when we multiply the likelihood function $L(Z_{k,j}^{(2)};\pi_k)$ by any constant $m$, a new term $m^{\delta_j}$ with the random variable $\delta_j$ will appear in the formulation of MPP, which leads to the violation of likelihood principle by changing the prior distribution of $\delta_j$. \textcolor{red}{(You can see if this explanation is more understandable)}

We assume that $Z_k^{(1)}$ and $Z_{k,j}^{(2)}$ are distributed as Binomial$(n_k^{(1)}, \pi_k)$ and Binomial$(n_{k,j}^{(2)}, \pi_k)$, respectively. The initial prior distributions $p_0(\pi_k)$ and $p_0(\delta_j)$ are Beta$(a_{\pi},b_{\pi})$ and Beta$(a_{\delta},b_{\delta})$, respectively. After plugging in these distributions and likelihood functions to Equation (\ref{eq:MPP}), we can analytically derive the MPP as follows, which is a multi-parameter version of the formula derived in \citealp{banbeta2019modified}:

%\begin{equation}
%\label{eq:MPP_snSMART}
%p_{MPP}(\boldsymbol{\pi},\boldsymbol{\delta}|\mathbf{D^{(2)}})\propto
%\frac{\prod_k\left\{\pi_k^{\sum_{j\in\Omega_k}Z_{k,j}^{(2)}\delta_j +a_{\pi}-1}(1-\pi_k)^{\sum_{j\in\Omega_k}\left(n_{k,j}^{(2)}-Z_{k,j}^{(2)}\right)\delta_j+b_{\pi}-1}\right\}\left\{\prod_j\frac{\delta_j^{a_{\delta}-1}(1-\delta_j)^{b_{\delta}-1}}{B(a_\delta,b_\delta)}\right\}}{\prod_{k}\left\{B\left(\sum_{j\in\Omega_k}Z_{k,j}^{(2)}\delta_j +a_{\pi},\sum_{j\in\Omega_k}\left(n_{k,j}^{(2)}-Z_{k,j}^{(2)}\right)\delta_j+b_{\pi}\right)\right\}}
%\end{equation}

\begin{equation}
\label{eq:MPP_snSMART}
p_{MPP}(\boldsymbol{\pi},\boldsymbol{\delta}|\mathbf{D^{(2)}})\propto
\frac{\prod_k\left\{\pi_k^{\sum_{j}Z_{k,j}^{(2)}\delta_j +a_{\pi}-1}(1-\pi_k)^{\sum_{j}\left(n_{k,j}^{(2)}-Z_{k,j}^{(2)}\right)\delta_j+b_{\pi}-1}\right\}\left\{\prod_j\frac{\delta_j^{a_{\delta}-1}(1-\delta_j)^{b_{\delta}-1}}{B(a_\delta,b_\delta)}\right\}}{\prod_{k}\left\{B\left(\sum_{j}Z_{k,j}^{(2)}\delta_j +a_{\pi},\sum_{j}\left(n_{k,j}^{(2)}-Z_{k,j}^{(2)}\right)\delta_j+b_{\pi}\right)\right\}}
\end{equation}

The choice of hyperparameters $a_\pi$, $b_\pi$, $a_\delta$ and $b_\delta$ reflects our belief in the response rates of treatments and the compatibility of ``current" and ``historical" in our snSMART. If we do not have previous knowledge about $\pi_k$ and $\delta_j$, their prior distribution can be set as Beta(1,1).

\subsection{Power prior model with closeness measure}
\label{sec:BM_FET}
In addition to likelihood-based approaches, we can define a metric that describes the closeness of the posterior distributions of first stage and second stage response rates. A natural choice of such a metric is Bhattacharyya's overlap measure (BOM) \citep{bhattacharyya1946measure}. If distributions from two populations are continuous with probability density functions $f_1(\theta)$ and $f_2(\theta)$, the BOM is defined as $\text{O}(f_1, f_2)=\int_{-\infty}^{\infty}\sqrt{f_1(\theta)f_2(\theta)}d\theta$. The BOM is useful in our setting because it takes values in the interval $[0,1]$, in which $\text{O}(f_1,f_2)=0$ indicates that two distributions are fully separated, while $\text{O}(f_1,f_2)=1$ means that two distributions are identical. This agrees with the interpretation of power parameters in power prior models.

We define the posterior distributions of response rates of treatment $k$ in stage 1 and stage 2 (within a specific subgroup $j$) as $p_1(\pi_k|\mathbf{D^{(1)}})$ and $p_{2j}(\pi_k|\mathbf{D^{(2)}})$, respectively, where $p_1(\pi_k|\mathbf{D^{(1)}})\propto L(Z_k^{(1)};\pi_k)p_0(\pi_k)$ and $p_{2j}(\pi_k|\mathbf{D^{(2)}})\propto L(Z_{k,j}^{(2)};\pi_k)p_0(\pi_k)$. Because we assume that the prior distributions of $\pi_k$ are Beta distributions, the posterior distributions $p_1$ and $p_{2j}$ will also follow Beta($a_1,b_1$) and Beta($a_{2j},b_{2j}$), respectively. Thus, we have
\begin{align}
\label{eq:BM}
\text{O}_k(p_1,p_{2j}) &= \int_0^1\sqrt{p_1(\pi_k|\mathbf{D^{(1)}})p_{2j}(\pi_k|\mathbf{D^{(2)}})}d\pi_k \nonumber \\
&= \int_0^1\sqrt{\frac{\pi_k^{a_1+a_{2j}-2}(1-\pi_k)^{b_1+b_{2j}-2}}{B(a_1,b_1)B(a_{2j},b_{2j})}}d\pi_k \nonumber \\
&= \frac{B(\frac{a_1+a_{2j}}{2},\frac{b_1+b_{2j}}{2})}{\sqrt{B(a_1,b_1)B(a_{2j},b_{2j})}}
\end{align}
where $a_1=Z_k^{(1)}+a_\pi$, $b_1=n_k^{(1)}-Z_k^{(1)}+b_\pi$, $a_{2j}=Z_{k,j}^{(2)}+a_\pi$, $b_{2j}=n_{k,j}^{(2)}-Z_{k,j}^{(2)}+b_\pi$. We then derive values for $\delta_1$ and $\delta_2$ as the average of BOM for all three treatments, or $\delta_j^{BOM}=\sum_k \text{O}_k(p_1,p_{2j})/3$.

Alternatively, the two-sided p-value of a Fisher's exact test (FET) from stage 1 and stage 2 data from subgroup $j$ can be used to quantify the closeness of treatment response rates in both stages. Specifically, we construct a $2\times 2$ table where the rows contain the numbers of participants from stage 1 or stage 2 subgroup $j$ and the columns contain the numbers of responders or non-responders. The two-sided p-value is computed using all the tables that are equally or more extreme than the observed table where extremity is defined by a table's hypergeometric probability. If the response rates change across the stages, the p-value from the FET should be small, suggesting that the data from stage 1 and stage 2 subgroup $j$ are incompatible. On the contrary, if the response rates do not change across the stages, we can expect a p-value close to 1, indicating that a higher weight should be put on the ``historical" data in subgroup $j$. Similar to the $\delta_j^{BOM}$, we can calculate $\delta_j^{FET}=\sum_k P_{k,j}/3$, in which $P_{k,j}$ is the p-value for subgroup $j$ and treatment $k$. 

\section{Simulation studies}
\label{sec:simulation}

\subsection{Data generation}
\label{sec:data}
We conducted Monte Carlo simulations to compare the performance of the power prior models described in Section \ref{sec:method}. The seven scenarios that we examined are listed in Table \ref{tab:scenario}. In all scenarios in stage 1, exactly 1/3 of participants are assigned to each of the three possible treatments. Their stage 1 responses are generated by a Bernoulli distribution with the response rates corresponding to the assigned treatments, shown in Table \ref{tab:scenario}(a). Their stage 2 responses are also generated by a Bernoulli distribution with the response rates corresponding to the assigned stage 1 and 2 treatments, shown in Table \ref{tab:scenario}(b). In scenarios 1-5, the first stage response rates of the treatments differ from each other, whereas these response rates are identical in scenarios 6 and 7. The last two scenarios can be used to examine the performance of estimation under the ``null" cases. 

The rationale of designing the scenarios is as follows:

\begin{description}
	\item[Scenario 1] The response rates remain unchanged in stage 2; there is full compatability between stage 1 and 2 data.
	\item[Scenario 2] The stage 2 response rates double if participants respond in stage 1; there is full compatibility between stage 1 data and stage 2 data only for stage 1 non-responders.
	\item[Scenario 3] The stage 2 response rates are halved for participants who do not respond in stage 1; there is full compatability between stage 1 data and stage 2 data only for stage 1 responders.
	\item[Scenario 4] The stage 2 response rates increase, but the scale of increase differs between stage 1 responders and non-responders; there is not full compatibility between stage 1 and stage 2 data.
	\item[Scenario 5] Stage 2 response rates change with respect to both first and second stage treatments, which violates a main assumption of the BJSM; there is not full compatibility between stage 1 and stage 2 data.
	\item[Scenario 6] All stage 1 and stage 2 response rates are equal; there is full compatibility between stage 1 and 2 data.
	\item[Scenario 7] Response rates are the same in stage 1 but not stage 2, and these depend on both first and second stage treatment (this violates a main assumption of the BJSM); there is not full compatibility between stage 1 and stage 2 data.
\end{description}

In Section \ref{sec:weight}, we use scenarios 1-4 to investigate the impact on $\delta_j$ when a part of or the whole stage 2 data are not compatible with the stage 1 data. We expect that: (1) both $\delta_1$ and $\delta_2$ are close to 1 in scenario 1; (2) $\delta_1$ should should move closer to 0 in scenario 2; (3) $\delta_2$ should move closer to 0 in scenario 3; (4) both $\delta_1$ and $\delta_2$ should move closer to 0 in scenarios 4. In Section \ref{sec:estimation}, we evaluate the estimation of $\pi_k$ using scenarios 4-7, with which we compare the performance either within different power prior models or between power prior models and the BJSM. We also examine whether partial borrowing of information from second stage data ($0<\delta_j<1$) can outperform situations when instead complete borrowing ($\delta_j=1$) or no borrowing ($\delta_j=0$) is applied.

The prior distribution of $\pi_k$ was Beta(1,1) for all methods, and the prior distribution of $\delta_j$ was Beta(1,1) in MPP. To maximize the flexibility of the BJSM, we set the prior distributions of both linkage parameters to gamma distributions with the support of $(0,\infty)$ and the prior mean of 1. All simulation studies were performed with 10,000 runs, and the total sample size for each run was either 90 or 300. 

\subsection{Estimation of $\delta_1$ and $\delta_2$}
\label{sec:weight}

In Table \ref{tab:bayes_delta}, we present the mean estimated $\delta_1$ and $\delta_2$ and their Monte Carlo standard errors obtained from five different power prior models in scenarios 1-4. Presently, we restrict our focus on scenarios 1-4 because these scenarios are designed to examine how \boldsymbol{$\delta$} changes when data from the two stages become incompatible.

When $N=90$, we first observe the differences in \boldsymbol{$\delta$} when comparing scenarios 2-4 to scenario 1, in which the data from stages 1 and 2 are fully compatible. In MLC, the mean estimated $\delta_1$ is 0.65 in scenario 1 compared to 0.32 in scenario 2 and 0.08 in scenario 4 where the stage 2 data from stage 1 responders is not compatible with the stage 1 data. The mean estimated $\delta_2$ is from 0.75 in scenario 1 compared to 0.40 in scenario 3 and 0.45 in scenario 4 where the stage 2 data from stage 1 non-responders is not compatible with the stage 1 data. 

Similarly, we can see the same pattern in MPP, PLC, BOM and FET, but the scale of difference varies. The differences are about 0.2 to 0.4 when comparing \boldsymbol{$\delta$} from scenario 1 to scenarios 2-4 in FET and BOM, 0.1 to 0.2 in MPP, and less than 0.1 in PLC. The differences become larger when $N=300$ for all methods. However, there is a trade-off between the difference in \boldsymbol{$\delta$} across various scenarios and the standard errors of estimated \boldsymbol{$\delta$}. The estimated \boldsymbol{$\delta$} from MLC have much larger standard errors than that of all other methods. In contrast, the estimates \boldsymbol{$\delta$} from PLC slightly change across different scenarios, resulting in relatively small standard errors of the estimates.

In addition, we also investigated the ranges of the mean estimated \boldsymbol{$\delta$} from different methods. When $N=90$, the values of \boldsymbol{$\delta$} from the BOM are close to 0.5 even when the data from two stages are incompatible, which indicates that the BOM tends to put higher weights on ``historical" data, regardless of the compatibility of first and second stages data. In contrast, the values of \boldsymbol{$\delta$} from PLC are between 0.2 and 0.35 in all scenarios, which agrees with the finding in \cite{ibrahim2003optimality} that the estimated \boldsymbol{$\delta$} from this method is relatively small in general. For MPP, MLC and FET, the values of \boldsymbol{$\delta$} are greater than 0.5 when data from two stages are compatible, whereas the values of \boldsymbol{$\delta$} are smaller than 0.5 if data are incompatible. 

We note that data compatibility is not the only driving force of the value of \boldsymbol{$\delta$} for MPP. The prior distribution of \boldsymbol{$\delta$} also plays an important role in the range of mean estimated \boldsymbol{$\delta$}. In Table \ref{tab:MPP_prior}, we let the prior distributions of \boldsymbol{$\delta$} be Beta(0.4,1.6), Beta(1,1), and Beta(1.6,0.4), which correspond to the prior means of 0.2, 0.5 and 0.8, respectively. We can see that the range of \boldsymbol{$\delta$} is centered at the prior mean of \boldsymbol{$\delta$}, especially when $N=90$. When $N=300$, the data have more capacity to adjust the estimated \boldsymbol{$\delta$} in addition to the influence from the prior distributions. Thus, we conclude, similar to \cite{neuenschwander2009note}, that the specification of the prior distribution of \boldsymbol{$\delta$} can greatly impact the results from the MPP method. The mean estimated \boldsymbol{$\delta$} under scenarios 5-7 for all methods can be found in the supplementary material Table \ref{suptab:bayes_delta_567}.

\begin{figure}[H]
	\centering
	\includegraphics[width=\textwidth]{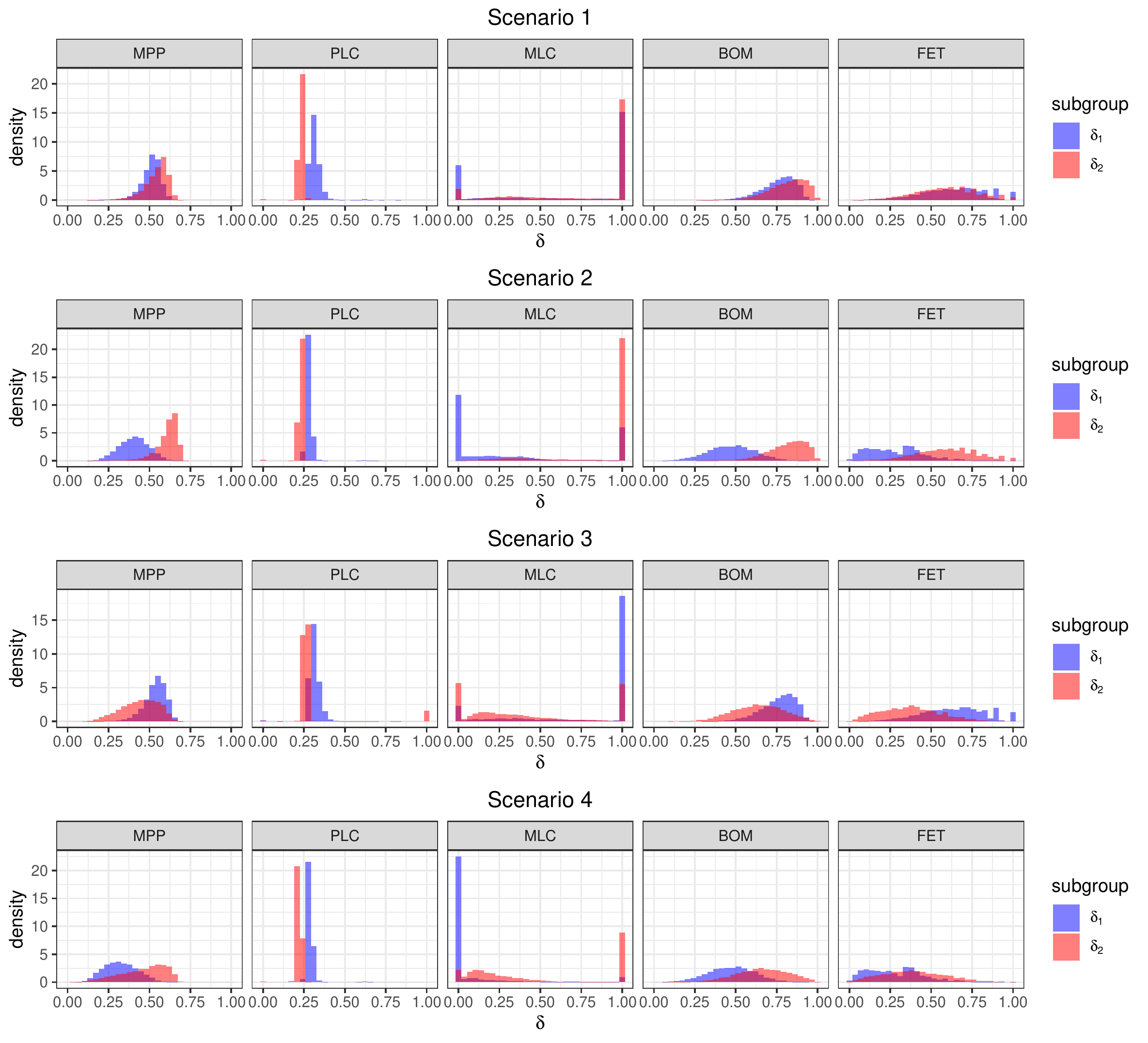}
	\caption{The distributions of $\delta_1$ and $\delta_2$ from modified power prior (MPP), power prior with penalized likelihood-type criterion (PLC), marginal likelihood criterion (MLC), Bhattacharyya's overlap measure (BOM) and measure from Fisher's exact test (FET) under scenarios 1-4. $N=90$}
	\label{fig:hist90}
\end{figure}

We further examine the distributions of estimated $\delta$ from different methods under scenarios 1-4 in Figure \ref{fig:hist90} when $N=90$. The histograms from the PLC under four scenarios do not differ much, indicating that the power parameters obtained from PLC do not vary with the changing scenarios. For MLC, the chance of choosing 0 or 1 for power parameters is extremely high, which is not a desirable property because second stage data are likely to be completely ignored even when the data across stages are fully compatible. This result suggests that the estimated $\delta$ from the power prior model with MLC is highly sensitive to slight changes in the number of responders. In particular, when the expected number of responders to a treatment in a subgroup in stage 2 is smaller than 10, which may be common in an snSMART, a change in the observed number of responders by 1 or 2 can result in a sharp decrease of the estimated $\delta$ from 1 to 0 or vice versa. The histograms from the MPP, BOM and FET are more appealing. In scenario 1, a large portion of distributions of $\delta_1$ and $\delta_2$ can overlap, while in other scenarios, we can easily see the move of either one or both distributions when part of or all the second stage data are not compatible with first stage data. 

When $N=300$, the distributions for $\delta_1$ and $\delta_2$ in supplementary Figure \ref{fig:hist300}. For FET, BOM and MPP, due to the increased sample size, the distributions move more when the data are incompatible, compared to the histograms in Figure \ref{fig:hist90}. For MLC, it seems that the chance of assigning the wrong power parameters becomes lower compared to $N=90$, but completely ignoring the second stage data is still undesirable even when the data across stages are not compatible. Borrowing some information from incompatible second stage data may still increase efficiency given that the bias may increase as well, which we will discuss in next Section \ref{sec:estimation}. The distributions of $\delta_1$ and $\delta_2$ under scenarios 5-7 can be found in the supplementary material Figures \ref{supfig:hist90_567} and \ref{supfig:hist300_567}.

\subsection{Estimation of $\pi$}
\label{sec:estimation}

\begin{figure}[h]
	\centering
	\includegraphics[width=\textwidth]{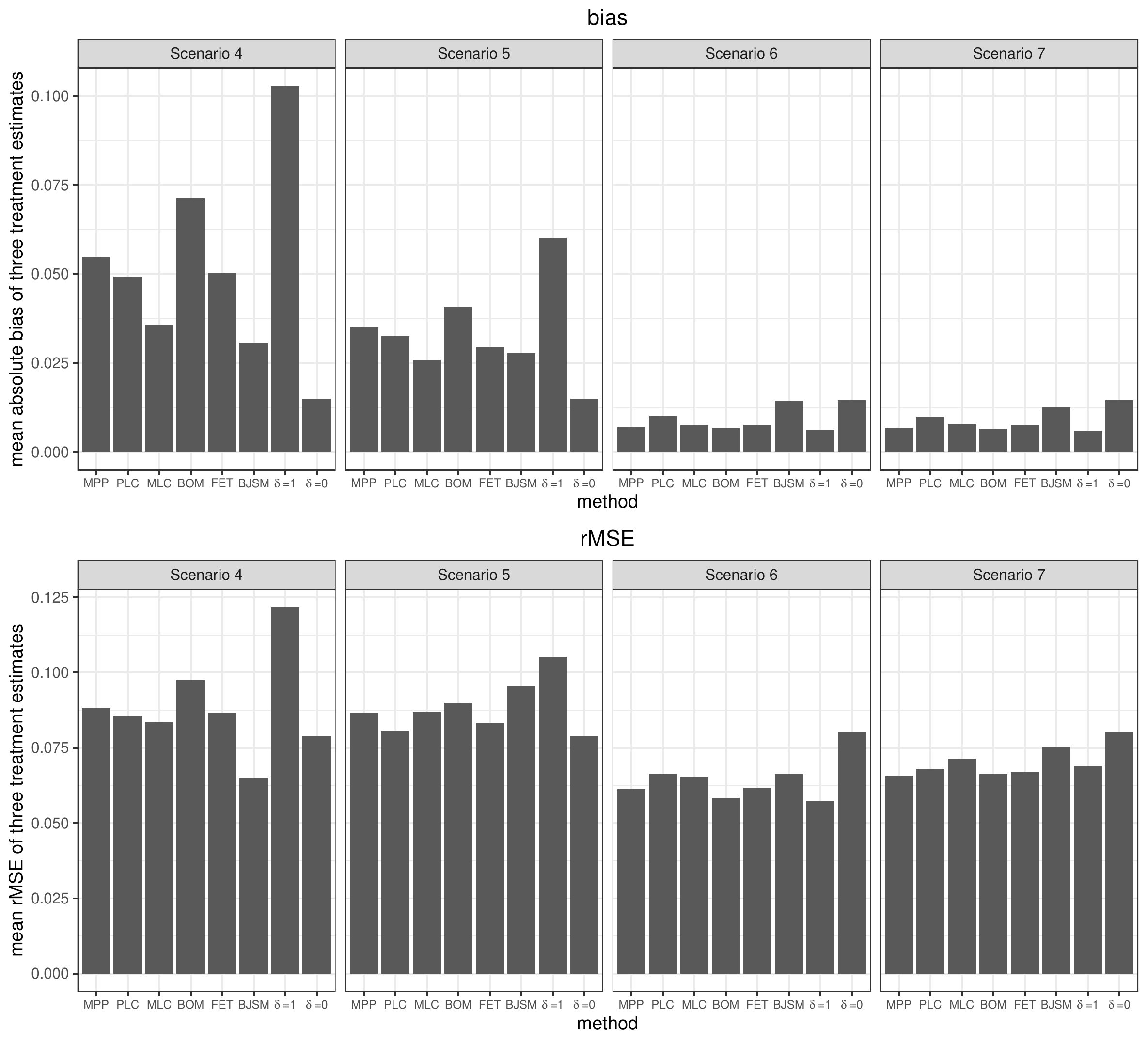}
	\caption{The barplots of the mean absolute biases and root mean squared errors (rMSEs) of the treatment response rate estimates under different methods. The results from scenarios 4-7 are shown. MPP=modified power prior model; PLC=power prior model with penalized likelihood-type criterion; MLC=power prior model with marginal likelihood criterion; BOM=power prior model with Bhattacharyya's overlap measure; FET=power prior models with Fisher's exact test; BJSM=Bayesian joint stage model. Power prior model is also applied with all $\delta$ fixed at 1 (or 0), meaning that the second stage data are completely used (or ignored). $N=90$}
	\label{fig:barplot}
\end{figure}

In Figure \ref{fig:barplot}, each bar is the simulation-wide average absolute value of bias or root mean squared error (rMSE) of the three treatment response rate estimates from each of the methods.  We include results for power prior models when \boldsymbol{$\delta$} is fixed at 0 or 1 for reference, as these two approaches only perform well in either fully compatible or highly incompatible scenarios, and are not preferred in most realistic settings. 

In scenario 4, we first note that BJSM has smallest bias and rMSE among all methods because the assumption of the linkage parameters is met in this scenario.  Among all the power prior methods,  we expect some bias because stage 2 data are highly incompatible with stage 1 data.  Although the estimation from MLC is least biased because the estimated \boldsymbol{$\delta$} are close to 0 in a large portion of simulated runs, we see that the rMSE of MLC is close to that from MPP, PLC and FET due to the high Monte Carlo variability of the MLC estimates. In scenario 5, the power prior models are more able to appropriately weight the second stage data, leading to lower rMSE compared to the BJSM because of violation to assumptions needed for the BJSM. 

In scenario 6, the data from two stages are compatible, and although the bias for all methods is small, we see that the rMSEs of BOM are smaller than other methods. This is because the distributions of $\delta_1$ and $\delta_2$ for BOM in supplementary Figure \ref{supfig:hist90_567} are clustered at the right half of the distribution, indicating power parameters closer to 1 compared to other histograms.  

Scenario 7 is similar to scenario 5 in terms of data incompatibility and violation of an assumption of the BJSM, but the level of data incompatibility is less strong according to Figure \ref{supfig:hist90_567}. Thus, we see that the rMSE of the power prior models is lower than the rMSE from the BJSM. The details of the bias and rMSE for all methods under scenarios 1-7 can be found in Tables \ref{suptab:bias_90} and \ref{suptab:rMSE_90} of the supplementary material.
We also have examined the patterns of bias and rMSE when $N = 60,75$ or $300$, and the patterns are similar to $N=90$ (results not shown). Thus, the power prior models can still be applied to snSMARTs with even smaller sample sizes.

\section{Discussion}
\label{sec:discussion}

%In addition, we summarized some existing power prior approaches and proposed new methods to choose the power parameters in the power prior models using the closeness measures, and we compared our methods to the existing power prior models and the BJSM. The power prior models were already widely used in the clinical trial studies when historical data were available. Previous researches focused more on investigating some operating characteristics of the trials with power prior models, like the power or the type I error rate \citep{hobbs2011hierarchical, ibrahim2015power, gravestock2017adaptive, pan2017calibrated,nikolakopoulos2018dynamic}, biases and/or mean squared errors of estimation \citep{ibrahim2015power,gravestock2017adaptive} and maximum tolerated doses in phase I trial design \citep{li2020pa}. However, investigation on the operating characteristics of the power parameters was often ignored. We thoroughly examined the distributions of the estimated power parameters in simulation studies, so we can better understand which methods are inappropriate in an snSMART.

Overall, we do not recommend use of the power prior models with the MPP, PLC or MLC. For the MPP, the choice of \boldsymbol{$\delta$} highly depends on its prior distribution, especially in an snSMART where the sample size is small. For the PLC, the estimated $\delta$ stay relatively constant across different scenarios in snSMARTs, regardless of whether the data from the two stages are compatible. For the MLC, the mean estimated \boldsymbol{$\delta$} can change along with the compatibility across two stages, but the value for \boldsymbol{$\delta$} is highly sensitive to small changes in number of responders in an snSMART. This sensitivity of the MLC leads to a high chance of choosing 0 or 1. 

Therefore, we feel that PLC and MLC should not be used to estimate response rates in an snSMART because it is undesirable to choose a fixed value or extreme values of 0 or 1 with high probability. MPP is not preferred as well because the weights highly depend on their prior distributions. 

Hence, the suggested candidate models for treatment effect estimation in an snSMART are the BJSM and the power prior models with BOM or FET when considering both the performance of treatment effect estimation and reasonable values of \boldsymbol{$\delta$}.

%We realize that when we conduct FET to decide the closeness between stage 1 outcomes and stage 2 outcomes from responders, both stage 1 and stage 2 outcomes from stage 1 responders are used. These repeated outcomes from stage 1 responders are likely positively correlated, which may lead to smaller p-values for $\delta_1$ because stage 1 responders are more likely to respond again in stage 2 if they receive the same treatment. However, we believe that the smaller p-values still meet our goal of choosing $\delta_1$ because a smaller weight should be put on the stage 2 outcomes from stage 1 responders when the within-individual correlation of outcomes can bias the estimated stage 2 response rates of treatments.

For the FET, we acknowledge that the stage 2 outcomes from stage 1 responders and their stage 1 outcomes may not be independent, which is an assumption of the Fisher's exact test, but we believe that the smaller p-values are reasonable because less weight should be put on second stage data when the within-individual correlation of first and second stage responses can affect the determination of dependency between stages and responses to treatment. Moreover, the number of correlated observations is likely to be small especially in rare disease trials. For the BOM, we can see that the assigned weights to ``historical" data tend to be larger than the weights from other methods. For the BJSM, we need to assume that the relationship between first and second stage response rates can be described proportionally through the linkage parameters. This assumption may be difficult to justify. 

When selecting a primary method of analysis, some background information about the treatments of interest in an snSMART may influence model choice in the estimation of treatment effects. If investigators believe that the second stage response rates are proportional to the first stage response rates and the proportionality (linkage) parameters do not depend on first and second stage treatments, then the BJSM may be preferred since it is most efficient when its assumptions are met. For example, the BJSM can be used if we believe that the response rates of all treatments will double in the second stage for all first stage responders. However, if this assumption is violated, which may be very likely, then power prior models with BOM or FET may be considered. The BOM is preferred if the data from two stages are more compatible, while the FET is preferred in the cases of less compatibility between data from two stages. If prior information about possible first and second stage response rates of all treatments exists, simulation studies can be conducted to help decide the prior distribution of \boldsymbol{$\delta$} and \boldsymbol{$\pi$}. 

An extension of the SMART is the proposal by \cite{liu2017sequential} that the design be enriched at later stages of the trial by the inclusion of subjects that received previous stage treatments outside of the trial. They used the term, SMARTER, for a SMART with enrichment. While this design assumed larger sample sizes, the same idea can apply to an snSMART. In an snSMART, it is not clear how a subject's information outside of the trial should be incorporated by the BJSM.  However, this enrichment is not a problem for the power prior model methods since these methods do not link an individual subject's responses between stages. Thus, our power prior models might be more appropriate for SMARTER designs.  

Moreover, a different number of subgroups in stage 2 of an snSMART can be pre-specified instead of $J=2$ in our study. In simulations, we have tried $J=6$, where the $\delta$ can differ depending on the individuals' first stage responses and their stage 1 treatment assignments. However, due to the resulting small sample sizes in each subgroup, the extra power parameters did not improve the bias and efficiency of the estimation (results not shown). The application of a Bhattacharyya's overlap measure or Fisher's exact test in power prior models is not limited to our snSMART settings, but also can be used in more general cases when data from historical trials are used to facilitate the data analysis of a current clinical trial. In this setting, the potential issue of independence between samples no longer exists because patients from different trials should be uncorrelated.

%This is a list of things I am going to include in this section.
%\begin{enumerate}
%	\item Restate that we provide a new idea of using power prior model in an snSMART 
%	\item We summarize some existing methods to choose $\delta$ and provide a novel approach (use of closeness measure) to choose $\delta$ as well.
%	\item State that there is no universally best approach to estimate response rates from an snSMART.
%	\begin{enumerate}
%		\item High variability of estimated $\delta$ from MLC limits its application. When the sample size of each arm in stage 1 is larger than 1000, MLC will be less likely to choose extreme values for $\delta$.
%		\item PLC can't choose different $\delta$ under different scenarios.
%		\item Choice of $\delta$ using MPP highly depend on the prior of $\delta$, especially when sample size is small. 
%		\item No theoretical basis for power prior model with FET, but its ability to choose $\delta$ is more desirable. 
%		\item For BM, when data across stages are incompatible, the distribution of estimated $\delta$ tend to become uniform.
%		\item BJSM does not perform well when the assumption on linkage parameters is violated. 
%	\end{enumerate}
%	\item Then what suggestions that we should give to practitioners? Describe when should we use these models? Should we say that practitioners should get more background knowledge before the trial? Should they try more than one method to see if the estimations are similar?
%	\item Any other thought?
%\end{enumerate}

\section{Acknowledgment}
This work was supported through a Patient-Centered Outcomes Research Institute (PCORI) Award (ME-1507-31108).

\bibliography{main.bib}

\begin{table}[H]
	\centering
	\caption{\label{tab:scenario} The true first and second stage response rates for simulation scenarios 1-7. (a) The response rates of the treatments in the first stage, which is the response rates of the interest. (b) The response rates of the treatments in the second stage, which depend on the first stage treatment and whether an individual responds to it. According to the snSMART design in Figure \ref{fig:design}, responders to their first stage treatment continue with the same treatments in the second stage, and the response rates of which are highlighted in gray. The non-highlighted response rates correspond to those from first stage non-responders.}
	\begin{tabular}{cccc}
		\multicolumn{4}{c}{(a) First stage response rates}\\
		\toprule
		& A & B & C \\
		\midrule
		Scenario 1-5 & 0.2 & 0.3 & 0.4 \\
		Scenario 6-7 & 0.3 & 0.3 & 0.3 \\
		\bottomrule
	\end{tabular}

	\bigskip
	
	\begin{tabular}{cccccccc}
		\multicolumn{8}{c}{(b) Second stage response rates}\\
		\toprule
		& & \multicolumn{3}{c}{Stage 1 treatment} & \multicolumn{3}{c}{Stage 1 treatment} \\
		\cmidrule(lr){3-5}\cmidrule(lr){6-8}
		& & A & B & C & A & B & C \\
		\midrule
		& & \multicolumn{3}{c}{\textbf{Scenario 1}} & \multicolumn{3}{c}{\textbf{Scenario 2}} \\
		\cmidrule(lr){3-5}\cmidrule(lr){6-8}
		\multirowcell{3}{Stage 2 \\treatment} & A & \cellcolor{lightgray}0.2 & 0.2 & 0.2 & \cellcolor{lightgray}0.4 & 0.2 & 0.2 \\
		& B & 0.3 & \cellcolor{lightgray}0.3 & 0.3 & 0.3 & \cellcolor{lightgray}0.6 & 0.3 \\
		& C & 0.4 & 0.4 & \cellcolor{lightgray}0.4 & 0.4 & 0.4 & \cellcolor{lightgray}0.8 \\
		\midrule
		& & \multicolumn{3}{c}{\textbf{Scenario 3}} & \multicolumn{3}{c}{\textbf{Scenario 4}} \\
		\cmidrule(lr){3-5}\cmidrule(lr){6-8}
		\multirowcell{3}{Stage 2 \\treatment} & A & \cellcolor{lightgray}0.2  & 0.1 & 0.1  & \cellcolor{lightgray}0.4  & 0.3 & 0.3  \\
		& B & 0.15 & \cellcolor{lightgray}0.3 & 0.15 & 0.45 & \cellcolor{lightgray}0.6 & 0.45 \\
		& C & 0.2  & 0.2 & \cellcolor{lightgray}0.4  & 0.6  & 0.6 &\cellcolor{lightgray} 0.8  \\
		\midrule
		& & \multicolumn{3}{c}{\textbf{Scenario 5}} & \multicolumn{3}{c}{\textbf{Scenario 6}} \\
		\cmidrule(lr){3-5}\cmidrule(lr){6-8}
		\multirowcell{3}{Stage 2 \\treatment} & A & \cellcolor{lightgray}0.6 & 0.4 & 0.4 & \cellcolor{lightgray}0.3 & 0.3 & 0.3  \\
		& B & 0.6 & \cellcolor{lightgray}0.6 & 0.15 & 0.3 & \cellcolor{lightgray}0.3  & 0.3 \\
		& C & 0.2 & 0.2 & \cellcolor{lightgray}0.6 & 0.3 & 0.3  & \cellcolor{lightgray}0.3 \\
		\midrule
		& & \multicolumn{3}{c}{\textbf{Scenario 7}} & \multicolumn{3}{c}{} \\
		\cmidrule(lr){3-5}
		\multirowcell{3}{Stage 2 \\treatment} & A & \cellcolor{lightgray}0.2 & 0.2 & 0.2 &  &  & \\
		& B & 0.3 & \cellcolor{lightgray}0.3 & 0.3 &  &  &  \\
		& C & 0.4 & 0.4 & \cellcolor{lightgray}0.4 &  &  &  \\
		\bottomrule
	\end{tabular}
\end{table}

\begin{sidewaystable}[H]
	\centering
	\caption{\label{tab:bayes_delta} The means and standard errors (in parentheses) of $\delta_1$ and $\delta_2$ obtained from each of the three power prior approaches, which are modified power prior model (MPP), power prior model with penalized likelihood-type criterion (PLC), power prior model with marginal likelihood criterion (MLC), power prior model with Bhattacharyya's overlap measure and power prior models with Fisher's exact test. Scenarios 1-4 in Table \ref{tab:scenario} are used to evaluate how these $\delta$s change with different levels of compatibility between first and second stage data. All simulation studies are done at $N=90$ or $300$.}
	\begin{tabular}{ccccccccccc}
		\toprule
		& \multicolumn{10}{c}{$N=90$}                                                    \\
		\cmidrule(lr){2-11}
		\multirowcell{2}{Scenario }& \multicolumn{2}{c}{MPP} & \multicolumn{2}{c}{PLC} & \multicolumn{2}{c}{MLC} & \multicolumn{2}{c}{BOM} & \multicolumn{2}{c}{FET} \\
		\cmidrule(lr){2-3}\cmidrule(lr){4-5}\cmidrule(lr){6-7}\cmidrule(lr){8-9}\cmidrule(lr){10-11}
		& $\delta_1$ & $\delta_2$ & $\delta_1$ & $\delta_2$ & $\delta_1$ & $\delta_2$ & $\delta_1$ & $\delta_2$ & $\delta_1$ & $\delta_2$\\
		1 & 0.51(0.06) & 0.54(0.07) & 0.32(0.04) & 0.23(0.02) & 0.65(0.42) & 0.75(0.35) & 0.76(0.10) & 0.81(0.11) & 0.64(0.19) & 0.59(0.18) \\
		2 & 0.41(0.09) & 0.61(0.06) & 0.28(0.03) & 0.23(0.02) & 0.32(0.39) & 0.87(0.26) & 0.48(0.14) & 0.81(0.11) & 0.28(0.17) & 0.59(0.18) \\
		3 & 0.53(0.07) & 0.44(0.11) & 0.31(0.04) & 0.30(0.17) & 0.76(0.36) & 0.40(0.36) & 0.76(0.10) & 0.64(0.15) & 0.64(0.19) & 0.38(0.18) \\
		4 & 0.32(0.10) & 0.46(0.14) & 0.29(0.03) & 0.22(0.02) & 0.08(0.21) & 0.45(0.39) & 0.48(0.14) & 0.66(0.16) & 0.28(0.17) & 0.40(0.19) \\
		\midrule\midrule
		& \multicolumn{10}{c}{$N=300$}                                                   \\
		\cmidrule(lr){2-11}
		\multirowcell{2}{Scenario }& \multicolumn{2}{c}{MPP} & \multicolumn{2}{c}{PLC} & \multicolumn{2}{c}{MLC} & \multicolumn{2}{c}{BOM} & \multicolumn{2}{c}{FET} \\
		\cmidrule(lr){2-3}\cmidrule(lr){4-5}\cmidrule(lr){6-7}\cmidrule(lr){8-9}\cmidrule(lr){10-11}
		& $\delta_1$ & $\delta_2$ & $\delta_1$ & $\delta_2$ & $\delta_1$ & $\delta_2$ & $\delta_1$ & $\delta_2$ & $\delta_1$ & $\delta_2$\\
		1 & 0.52(0.06) & 0.55(0.07) & 0.23(0.10) & 0.13(0.04) & 0.67(0.42) & 0.77(0.34) & 0.75(0.11) & 0.81(0.12) & 0.57(0.18) & 0.55(0.18) \\
		2 & 0.25(0.10) & 0.66(0.06) & 0.18(0.01) & 0.14(0.01) & 0.13(0.22) & 0.87(0.25) & 0.20(0.11) & 0.81(0.12) & 0.08(0.09) & 0.55(0.18) \\
		3 & 0.60(0.07) & 0.26(0.11) & 0.20(0.02) & 0.16(0.01) & 0.87(0.26) & 0.14(0.17) & 0.75(0.11) & 0.35(0.15) & 0.57(0.18) & 0.13(0.12) \\
		4 & 0.10(0.04) & 0.28(0.15) & 0.18(0.01) & 0.13(0.02) & 0.00(0.01) & 0.15(0.18) & 0.20(0.11) & 0.41(0.16) & 0.08(0.09) & 0.17(0.14) \\
		\bottomrule
	\end{tabular}
	
\end{sidewaystable}

\begin{table}[H]
	\centering
	\caption{\label{tab:MPP_prior} The means and standard errors (in parentheses) of $\delta_1$ and $\delta_2$ obtained from modified power prior model (MPP) with different $E(\delta)$, or prior mean of $\delta$. Scenarios 1-4 in Table \ref{tab:scenario} are used to evaluate how these $\delta$s change with different levels of compatibility between first and second stage data. All simulation studies are done at $N=90$ or $300$.}
	\begin{tabular}{ccccccc}
		\toprule
		& \multicolumn{6}{c}{$N=90$}                                                                                           \\
		\cmidrule(lr){2-7}
		\multirow{2}{*}{Scenario} & \multicolumn{2}{c}{$E(\delta)=0.20$} & \multicolumn{2}{c}{$E(\delta)=0.50$} & \multicolumn{2}{c}{$E(\delta)=0.80$} \\
		\cmidrule(lr){2-3}\cmidrule(lr){4-5}\cmidrule(lr){6-7}
		& $\delta_1$        & $\delta_2$       & $\delta_1$        & $\delta_2$       & $\delta_1$        & $\delta_2$       \\
		1                         & 0.23(0.05)        & 0.31(0.06)       & 0.51(0.06)        & 0.54(0.07)       & 0.80(0.04)        & 0.81(0.05)       \\
		2                         & 0.15(0.05)        & 0.36(0.07)       & 0.41(0.09)        & 0.61(0.06)       & 0.73(0.08)        & 0.86(0.03)       \\
		3                         & 0.25(0.05)        & 0.23(0.08)       & 0.54(0.07)        & 0.44(0.11)       & 0.82(0.05)        & 0.73(0.11)       \\
		4                         & 0.11(0.05)        & 0.23(0.10)       & 0.33(0.10)        & 0.46(0.14)       & 0.65(0.11)        & 0.75(0.13)       \\
		\midrule\midrule 
		& \multicolumn{6}{c}{$N=300$} \\
		\cmidrule(lr){2-7}
		\multirow{2}{*}{Scenario} & \multicolumn{2}{c}{$E(\delta)=0.20$} & \multicolumn{2}{c}{$E(\delta)=0.50$} & \multicolumn{2}{c}{$E(\delta)=0.80$} \\
		\cmidrule(lr){2-3}\cmidrule(lr){4-5}\cmidrule(lr){6-7}
		& $\delta_1$        & $\delta_2$       & $\delta_1$        & $\delta_2$       & $\delta_1$        & $\delta_2$       \\
		1                         & 0.25(0.05)        & 0.33(0.07)       & 0.52(0.06)        & 0.55(0.08)       & 0.81(0.04)        & 0.81(0.05)       \\
		2                         & 0.08(0.04)        & 0.42(0.07)       & 0.25(0.10)        & 0.66(0.06)       & 0.50(0.15)        & 0.88(0.04)       \\
		3                         & 0.34(0.07)        & 0.12(0.06)       & 0.60(0.07)        & 0.26(0.11)       & 0.85(0.04)        & 0.49(0.17)       \\
		4                         & 0.03(0.02)        & 0.14(0.09)       & 0.10(0.04)        & 0.28(0.15)       & 0.22(0.11)        & 0.51(0.23)      \\
		\bottomrule
	\end{tabular}
\end{table}

\end{document}

\setlength{\textheight}{625pt} \setlength{\baselineskip}{26pt}	
\title{Supplementary Materials to ``Power Prior Models for Treatment Effect Estimation in a Small n, Sequential, Multiple Assignment, Randomized Trial"}
\author[1]{Yan-Cheng Chao}
\author[1]{Thomas M. Braun}
\author[2]{Roy N. Tamura}
\author[1]{Kelley M. Kidwell}
\affil[1]{\footnotesize Department of Biostatistics, School of Public Health, University of Michigan, Ann Arbor, MI 48109, U.S.A.}
\affil[2]{\footnotesize Health Informatics Institute, University of South Florida, Tampa, FL 33620, U.S.A.}
\date{}
\maketitle

\renewcommand{\thetable}{S\arabic{table}}   
\renewcommand{\thefigure}{S\arabic{figure}}

\begin{sidewaystable}[h]
	\centering
	\caption{\label{suptab:bayes_delta_567} The means and standard errors (in parentheses) of $\delta_1$ and $\delta_2$ obtained from each of the three power prior approaches, which are modified power prior model (MPP), power prior model with penalized likelihood-type criterion (PLC), and power prior model with marginal likelihood criterion (MLC). Scenarios 5-7 in Table \ref{tab:scenario} are used to evaluate how these $\delta$s change with different levels of compatibility between first and second stage data. All simulation studies are done at $N=90$ or $300$.}
	\begin{tabular}{ccccccccccc}
		\toprule
		& \multicolumn{10}{c}{$N=90$}                                                    \\
		\cmidrule(lr){2-11}
		\multirowcell{2}{Scenario }& \multicolumn{2}{c}{MPP} & \multicolumn{2}{c}{PLC} & \multicolumn{2}{c}{MLC} & \multicolumn{2}{c}{BOM} & \multicolumn{2}{c}{FET} \\
		\cmidrule(lr){2-3}\cmidrule(lr){4-5}\cmidrule(lr){6-7}\cmidrule(lr){8-9}\cmidrule(lr){10-11}
		& $\delta_1$ & $\delta_2$ & $\delta_1$ & $\delta_2$ & $\delta_1$ & $\delta_2$ & $\delta_1$ & $\delta_2$ & $\delta_1$ & $\delta_2$\\
		5 & 0.42(0.10) & 0.47(0.13) & 0.28(0.03) & 0.22(0.02) & 0.29(0.37) & 0.42(0.37) & 0.51(0.15) & 0.61(0.17) & 0.30(0.18) & 0.35(0.19) \\
		6 & 0.51(0.06) & 0.54(0.07) & 0.32(0.04) & 0.23(0.02) & 0.65(0.42) & 0.75(0.35) & 0.76(0.11) & 0.81(0.11) & 0.62(0.19) & 0.59(0.18) \\
		7 & 0.49(0.07) & 0.50(0.10) & 0.32(0.04) & 0.23(0.02) & 0.56(0.44) & 0.62(0.40) & 0.74(0.11) & 0.76(0.13) & 0.59(0.20) & 0.52(0.19) \\
		\midrule\midrule
		& \multicolumn{10}{c}{$N=300$}                                                   \\
		\cmidrule(lr){2-11}
		\multirowcell{2}{Scenario }& \multicolumn{2}{c}{MPP} & \multicolumn{2}{c}{PLC} & \multicolumn{2}{c}{MLC} & \multicolumn{2}{c}{BOM} & \multicolumn{2}{c}{FET} \\
		\cmidrule(lr){2-3}\cmidrule(lr){4-5}\cmidrule(lr){6-7}\cmidrule(lr){8-9}\cmidrule(lr){10-11}
		& $\delta_1$ & $\delta_2$ & $\delta_1$ & $\delta_2$ & $\delta_1$ & $\delta_2$ & $\delta_1$ & $\delta_2$ & $\delta_1$ & $\delta_2$\\
		5 & 0.21(0.10) & 0.27(0.14) & 0.17(0.01) & 0.14(0.01) & 0.04(0.08) & 0.10(0.11) & 0.21(0.11) & 0.33(0.15) & 0.06(0.08) & 0.13(0.12) \\
		6 & 0.52(0.06) & 0.55(0.07) & 0.24(0.11) & 0.13(0.04) & 0.67(0.41) & 0.77(0.34) & 0.75(0.11) & 0.81(0.12) & 0.57(0.18) & 0.55(0.18) \\
		7 & 0.46(0.10) & 0.44(0.12) & 0.24(0.11) & 0.13(0.04) & 0.44(0.43) & 0.43(0.40) & 0.67(0.13) & 0.66(0.15) & 0.46(0.19) & 0.38(0.18) \\
		\bottomrule
	\end{tabular}
	
\end{sidewaystable}

\begin{figure}[h]
	\centering
	\includegraphics[width=\textwidth]{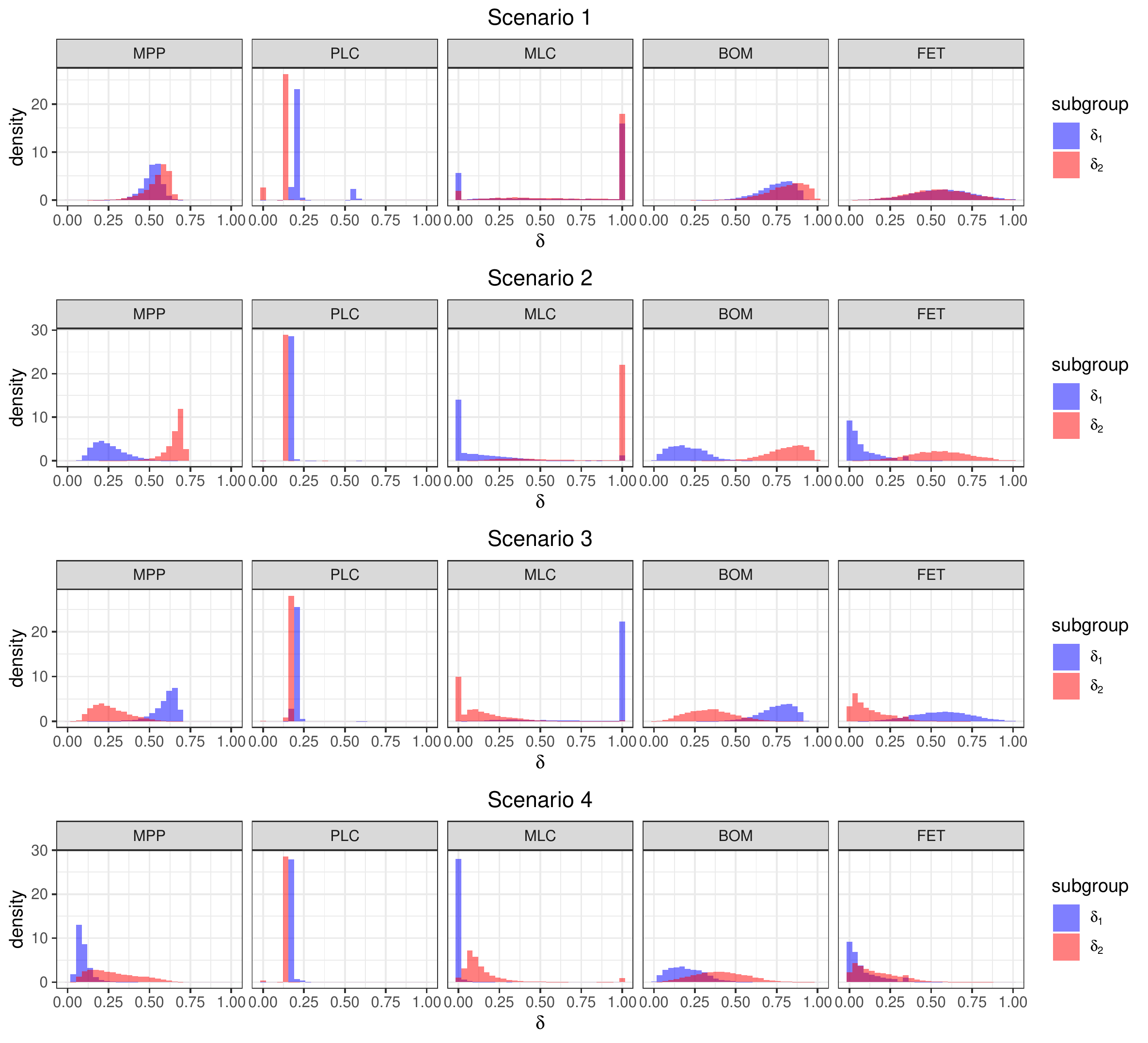}
	\caption{The distributions of $\delta_1$ and $\delta_2$ from modified power prior (MPP), power prior with penalized likelihood-type criterion (PLC), marginal likelihood criterion (MLC), Bhattacharyya's overlap measure (BOM) and measure from Fisher's exact test (FET) under scenarios 1-4. $N=300$}
	\label{fig:hist300}
\end{figure}

\begin{figure}[h]
	\centering
	\includegraphics[width=\textwidth]{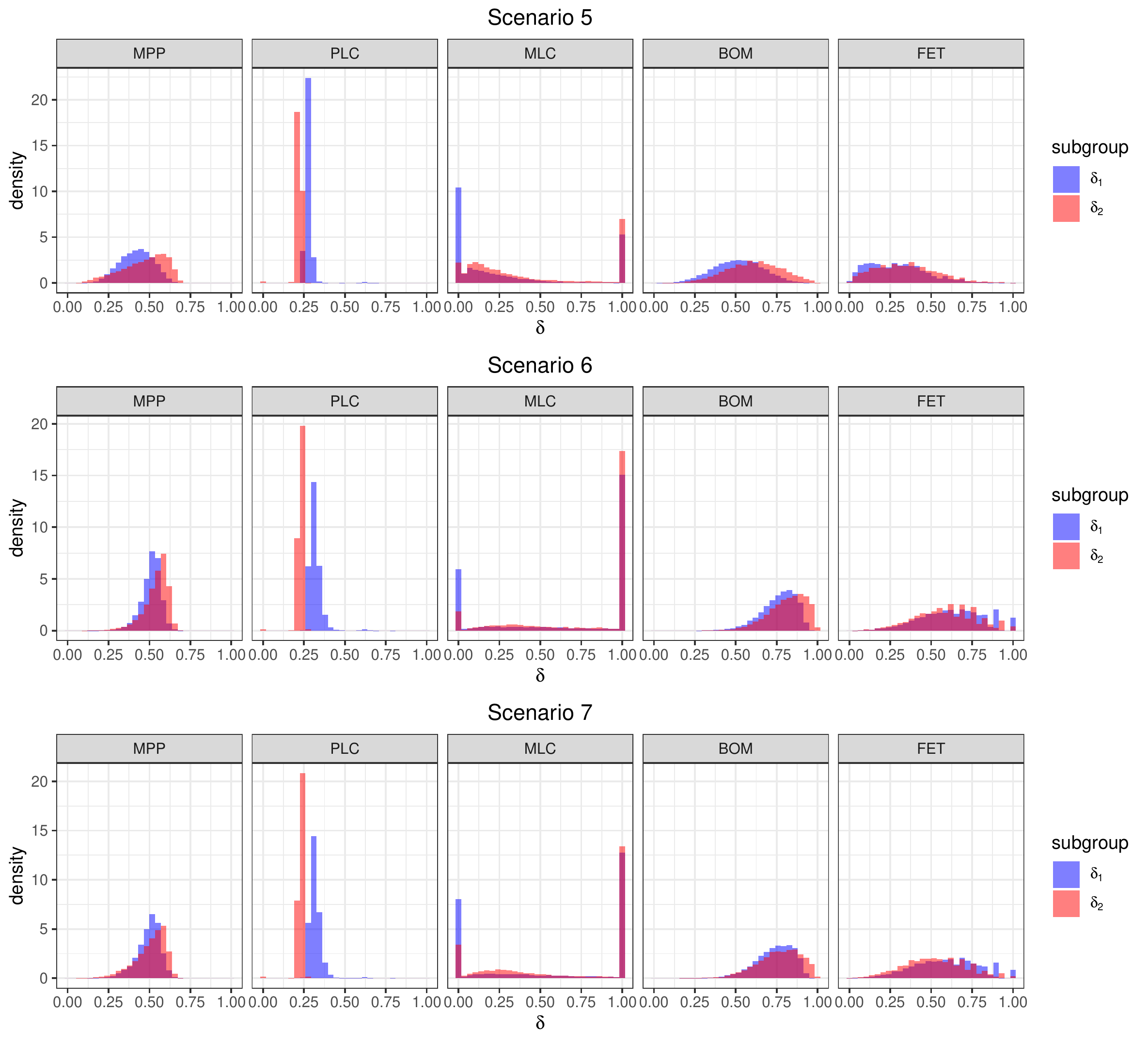}
	\caption{The distributions of $\delta_1$ and $\delta_2$ from modified power prior (MPP), power prior with penalized likelihood-type criterion (PLC), marginal likelihood criterion (MLC), Bhattacharyya's overlap measure (BOM) and measure from Fisher's exact test (FET) under scenarios 5-7. $N=90$}
	\label{supfig:hist90_567}
\end{figure}

\begin{figure}[h]
	\centering
	\includegraphics[width=\textwidth]{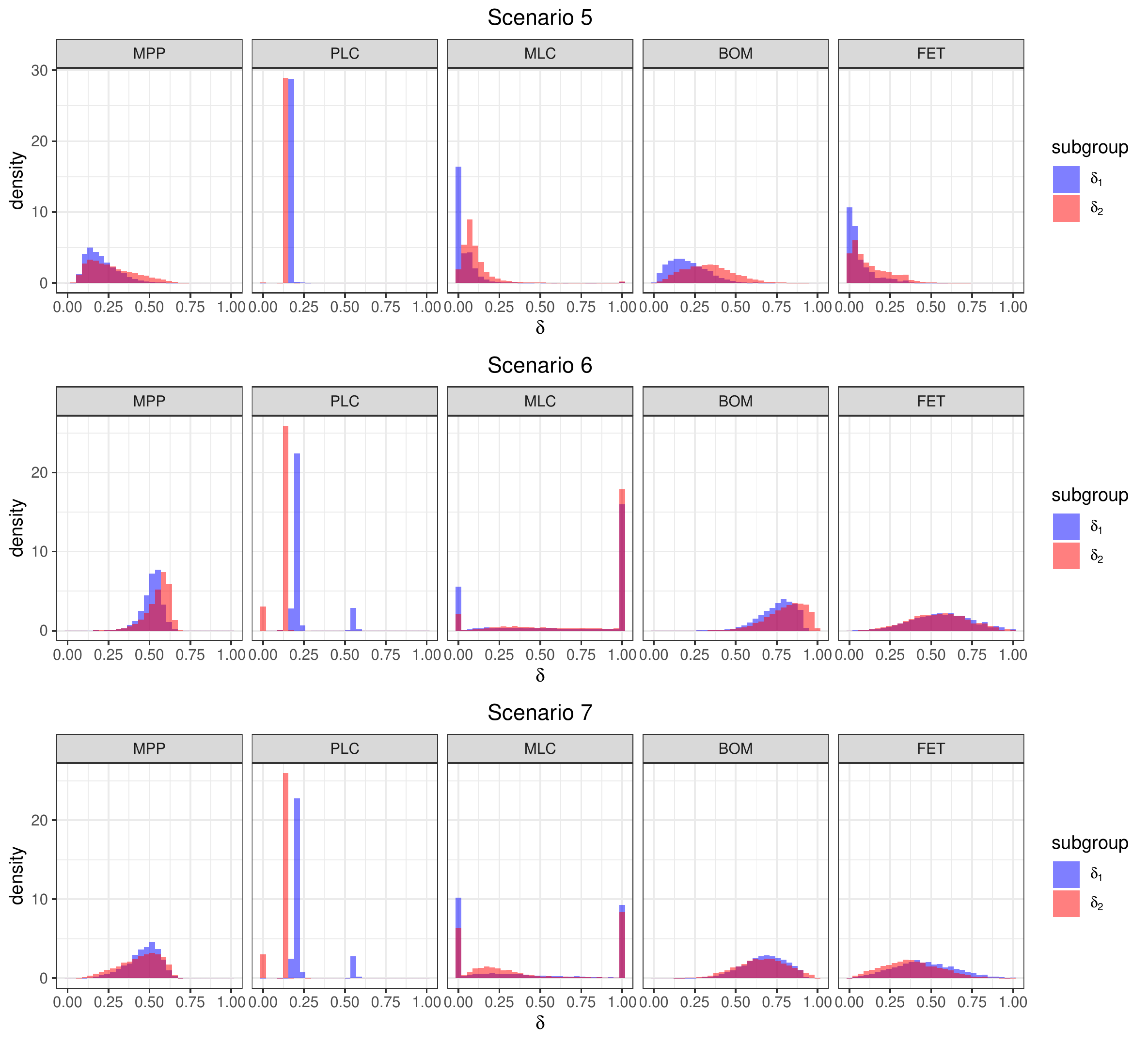}
	\caption{The distributions of $\delta_1$ and $\delta_2$ from modified power prior (MPP), power prior with penalized likelihood-type criterion (PLC), marginal likelihood criterion (MLC), Bhattacharyya's overlap measure (BOM) and measure from Fisher's exact test (FET) under scenarios 5-7. $N=300$}
	\label{supfig:hist300_567}
\end{figure}

\begin{table}[h]
	\centering
	\caption{\label{suptab:bias_90} The bias of the estimates of treatment response rates under different methods. MPP, PLC, MLC, BOM, FET and BJSM stand for modified power prior model, power prior model with penalized likelihood-type criterion, power prior model with marginal likelihood criterion, power prior model with Bhattacharyya's overlap measure and power prior models with Fisher's exact test and Bayesian joint stage model, respectively. Power prior model is also applied with all $\delta$ fixed at 1 (or 0), meaning that the second stage data are completely used (or ignored). $N=90$.}
	\begin{tabular}{cccccccccc}
		\toprule
		Scenario & Treatment & MPP & PLC & MLC & BOM & FET & BJSM & $\delta=1$ & $\delta=0$ \\
		\midrule
		\multirow{3}{*}{1} & A & 0.012  & 0.015  & 0.011  & 0.010  & 0.011  & 0.015  & 0.010  & 0.020 \\
		& B & 0.011  & 0.012  & 0.010  & 0.008  & 0.009  & 0.016  & 0.008  & 0.017 \\
		& C & -0.001 & 0.005  & 0.003  & 0.003  & 0.004  & 0.013  & 0.002  & 0.008 \\
		\midrule
		\multirow{3}{*}{2} & A & 0.023  & 0.024  & 0.018  & 0.021  & 0.018  & 0.019  & 0.030  & 0.020 \\
		& B & 0.032  & 0.032  & 0.024  & 0.031  & 0.024  & 0.025  & 0.051  & 0.017 \\
		& C & 0.036  & 0.038  & 0.027  & 0.044  & 0.032  & 0.032  & 0.075  & 0.008 \\
		\midrule
		\multirow{3}{*}{3} & A & -0.005 & 0.000  & 0.000  & -0.013 & -0.002 & 0.011  & -0.025 & 0.020 \\
		& B & -0.015 & -0.010 & -0.007 & -0.027 & -0.013 & 0.009  & -0.043 & 0.017 \\
		& C & -0.037 & -0.026 & -0.020 & -0.044 & -0.027 & 0.004  & -0.064 & 0.008 \\
		\midrule
		\multirow{3}{*}{4} & A & 0.039  & 0.036  & 0.030  & 0.046  & 0.034  & 0.022  & 0.065  & 0.020 \\
		& B & 0.057  & 0.050  & 0.037  & 0.070  & 0.049  & 0.031  & 0.102  & 0.017 \\
		& C & 0.068  & 0.062  & 0.040  & 0.097  & 0.068  & 0.039  & 0.141  & 0.008 \\
		\midrule
		\multirow{3}{*}{5} & A & 0.070  & 0.056  & 0.051  & 0.083  & 0.058  & 0.099  & 0.120  & 0.020 \\
		& B & 0.054  & 0.045  & 0.038  & 0.061  & 0.042  & 0.051  & 0.089  & 0.017 \\
		& C & -0.018 & -0.003 & -0.011 & -0.022 & -0.011 & -0.066 & -0.028 & 0.008 \\
		\midrule
		\multirow{3}{*}{6} & A & 0.007  & 0.009  & 0.006  & 0.006  & 0.007  & 0.014  & 0.005  & 0.013 \\
		& B & 0.010  & 0.011  & 0.009  & 0.007  & 0.008  & 0.015  & 0.007  & 0.016 \\
		& C & 0.004  & 0.010  & 0.007  & 0.007  & 0.008  & 0.014  & 0.006  & 0.014 \\
		\midrule
		\multirow{3}{*}{7} & A & -0.022 & -0.010 & -0.016 & -0.033 & -0.021 & -0.035 & -0.043 & 0.013 \\
		& B & 0.010  & 0.011  & 0.010  & 0.008  & 0.009  & 0.014  & 0.007  & 0.016 \\
		& C & 0.032  & 0.029  & 0.030  & 0.045  & 0.036  & 0.059  & 0.054  & 0.014 \\
		\bottomrule
	\end{tabular}
	
\end{table}

\begin{table}[h]
	\centering
	\caption{\label{suptab:rMSE_90} The root mean square error (rMSE) of the estimates of treatment response rates under different methods. MPP, PLC, MLC, BOM, FET and BJSM stand for modified power prior model, power prior model with penalized likelihood-type criterion, power prior model with marginal likelihood criterion, power prior model with Bhattacharyya's overlap measure, power prior models with Fisher's exact test and Bayesian joint stage model, respectively. Power prior model is also applied with all $\delta$ fixed at 1 (or 0), meaning that the second stage data are completely used (or ignored). $N=90$.}
	\begin{tabular}{cccccccccc}
		\toprule
		Scenario & Treatment & MPP & PLC & MLC & BOM & FET & BJSM & $\delta=1$ & $\delta=0$ \\
		\midrule
		\multirow{3}{*}{1} & A & 0.055 & 0.060 & 0.058 & 0.053 & 0.056 & 0.058 & 0.052 & 0.071 \\
		& B & 0.062 & 0.067 & 0.065 & 0.059 & 0.062 & 0.066 & 0.058 & 0.080 \\
		& C & 0.064 & 0.069 & 0.069 & 0.060 & 0.063 & 0.072 & 0.059 & 0.085 \\
		\midrule
		\multirow{3}{*}{2} & A & 0.061 & 0.066 & 0.061 & 0.061 & 0.061 & 0.057 & 0.065 & 0.071 \\
		& B & 0.073 & 0.077 & 0.072 & 0.072 & 0.072 & 0.065 & 0.083 & 0.080 \\
		& C & 0.080 & 0.085 & 0.082 & 0.082 & 0.078 & 0.074 & 0.102 & 0.085 \\
		\midrule
		\multirow{3}{*}{3} & A & 0.056 & 0.058 & 0.061 & 0.055 & 0.059 & 0.060 & 0.054 & 0.071 \\
		& B & 0.065 & 0.067 & 0.072 & 0.066 & 0.066 & 0.069 & 0.069 & 0.080 \\
		& C & 0.076 & 0.076 & 0.080 & 0.077 & 0.073 & 0.075 & 0.087 & 0.085 \\
		\midrule
		\multirow{3}{*}{4} & A & 0.072 & 0.071 & 0.071 & 0.075 & 0.071 & 0.055 & 0.087 & 0.071 \\
		& B & 0.090 & 0.086 & 0.085 & 0.097 & 0.087 & 0.064 & 0.121 & 0.080 \\
		& C & 0.103 & 0.099 & 0.095 & 0.120 & 0.102 & 0.075 & 0.156 & 0.085 \\
		\midrule
		\multirow{3}{*}{5} & A & 0.096 & 0.085 & 0.090 & 0.105 & 0.088 & 0.114 & 0.136 & 0.071 \\
		& B & 0.088 & 0.084 & 0.086 & 0.091 & 0.084 & 0.079 & 0.110 & 0.080 \\
		& C & 0.076 & 0.074 & 0.085 & 0.074 & 0.078 & 0.094 & 0.070 & 0.085 \\
		\midrule
		\multirow{3}{*}{6} & A & 0.061 & 0.066 & 0.065 & 0.058 & 0.062 & 0.066 & 0.057 & 0.079 \\
		& B & 0.062 & 0.067 & 0.065 & 0.059 & 0.062 & 0.067 & 0.058 & 0.080 \\
		& C & 0.061 & 0.067 & 0.066 & 0.058 & 0.061 & 0.066 & 0.058 & 0.081 \\
		\midrule
		\multirow{3}{*}{7} & A & 0.063 & 0.064 & 0.069 & 0.064 & 0.064 & 0.071 & 0.067 & 0.079 \\
		& B & 0.063 & 0.067 & 0.068 & 0.059 & 0.063 & 0.067 & 0.058 & 0.080 \\
		& C & 0.072 & 0.074 & 0.077 & 0.076 & 0.074 & 0.088 & 0.081 & 0.081 \\
		\bottomrule
	\end{tabular}
	
\end{table}